\documentclass[aps,prl,10pt,twocolumn,superscriptaddress,nofootinbib]{revtex4-2}
\usepackage{amsmath,amsfonts,amssymb,bbm, graphicx,color,tikz,pgf,svg, csquotes}
\usepackage[colorlinks=True, citecolor=jblue, linkcolor=jred, urlcolor=jblue]{hyperref}
\definecolor{jred}{rgb}{0.8,0,0}
\definecolor{jgreen}{rgb}{0,0.7,0}
\definecolor{jblue}{rgb}{0,0,0.8}
\tikzstyle{v} =  [circle, draw=black, line width=.2pt, fill=black, inner sep=0pt, minimum size=1mm]
\tikzstyle{wv} = [circle, inner sep=0.1pt, draw=jred, minimum size=1.5mm]
\tikzstyle{bv} = [circle, inner sep=0.1pt, fill=jred, minimum size=1.5mm]
\tikzstyle{e} =	[draw=jred,line width=1pt]
\tikzstyle{eb}= [draw=jgreen,line width=1pt]

\renewcommand{\eqref}[1]{Eq.~(\ref{#1})}
\renewcommand\[{\begin{equation}}
\renewcommand\]{\end{equation}}

\def\d{\mathrm{d}}

\newcommand{\Z}{\mathbb Z}

\renewcommand{\O}{\textrm{O}}
\newcommand{\U}{\textrm{U}}
\newcommand{\rk}{r}
\newcommand{\gm}{G}
\newcommand{\gv}{V}
\newcommand{\ak}{N_k}

\newcommand{\gf}{\varphi}
\newcommand{\gfb}{\bar\varphi}

\newcommand{\reg}{{\mathcal R}_k}
\newcommand{\wfr}{\bar{Z}}
\newcommand{\wfc}{{Z_c}}

\newcommand{\ks}{{2}}          
\renewcommand{\k}{k^{2}}
\newcommand{\zc}{{2}}

\newcommand{\ad}{\bar{\eta}}
\newcommand{\ec}{\eta_c}
\newcommand{\efac}{F(\ec)}
\newcommand{\efacs}{F_s(\ad)}
\newcommand{\efacsa}{F_{s_a}\!(\eta^a,\eta^b)}

\newcommand{\efd}{{d_\rk}}
\newcommand{\crd}{d_{\textrm{crit}}}
\newcommand{\nosed}{d_{\bullet}}

\newcommand{\m}{\mu}
\newcommand{\mr}{\tilde{\mu}}

\newcommand{\cnrc}{\tilde{\lambda}_{n}^c}
\newcommand{\cn}[1]{\lambda_{#1}}
\newcommand{\cnr}[1]{\tilde{\lambda}_{#1}}   
\newcommand{\bnc}[1]{\beta_{#1}(\mr,\cnr{i}^c)}
\newcommand{\bn}[1]{\beta_{#1}(\mr,\cnr{i})}
\newcommand{\cer}[1]{\theta_{#1}^\textsc{r}}
\newcommand{\cea}[1]{\theta_{#1}^\textsc{a}}
\newcommand{\ceg}[1]{\theta_{#1}^\textsc{g}}
\newcommand{\nmax}{n_\textrm{max}}

\begin{document}

\title{New Fixed Points from Melonic Interactions}

\author{Leonardo Juliano}
\email{leonardo.juliano@sns.it}
\affiliation{Scuola Normale Superiore, Piazza dei Cavalieri 7, 56126, Pisa, Italy}\affiliation{
INFN Sezione di Pisa, Largo Pontecorvo 3, 56127 Pisa, Italy}

\author{Johannes Th\"urigen}
\email{johannes.thuerigen@uni-muenster.de}
\affiliation{Mathematical Institute, University of M\"unster, Einsteinstr. 62, 48149 M\"unster, Germany}

\begin{abstract} 
Generalizations of vector field theories to tensors allow to similarly apply large-$N$ techniques but find a richer though often still tractable structure.
However, the potential of such tensor theories has not been fully exploited since only a symmetry-reduced ``isotropic'' 
part of their phase space has been studied so far. 
Here we present for the first time the richness of the tensorial phase space applying the functional renormalization group to tensor fields of rank $\rk$ in the cyclic-melonic potential approximation including the flow of anomalous dimensions. 
Due to a decoupling of the flow equations of the $\rk$ couplings at given order, we find non-Gaussian fixed points in any regime of $0<s\le\rk$ non-vanishing coupling types.
Each of these regimes contains isotropic fixed points of Wilson-Fisher type as well as new anisotropic fixed points.
This new classification reveals a rich structure of renormalization group dynamics 
including candidates for asymptotic safe fixed points even at critical dimension.
Considering the tensorial interactions as generating discrete geometries, the various fixed points correspond to continuum limits of distinguished ensembles of triangulations raising hope to find new classes of continuum geometry in this way. 
\end{abstract}

\maketitle


Generalizations of multi-scalar field theories \cite{Pelissetto:2002hz} from vector to tensor fields have attracted much attention in recent years, both as generating functions of lattice quantum gravity \cite{Rivasseau:1112, Reisenberger:0002b, Freidel:0505, Oriti:1110, Carrozza:13} and as conformal field theories related to Sachdev-Ye-Kitaev models \cite{Witten:1610, Klebanov:18, Benedetti:2004}.
However, they are usually studied in a symmetry-reduced regime where there is only a single coupling for all interactions of the same kind.
This leaves the question whether there could be new regimes of interest in the full theory space of such tensor theories.

Here, we address this question in the context of field theories with tensorial interactions \cite{BenGeloun:1111,BenGeloun:1201, BenGeloun:1306} which can be seen as higher-rank generalization of the Kontsevich model \cite{Kontsevich:1992en} and matrix field theory \cite{Grosse:0401, Wulkenhaar:2019, Hock:2020} and are referred to as (tensorial) group field theory in the quantum-gravity context \cite{Freidel:0505, Oriti:1110, Carrozza:13}.
These are specific in that the $\rk$ tensor arguments of the fields are the propagating degrees of freedom but they interact not point-like but tensorially.
That means the theory space consists of interactions with unitary symmetry in each argument. Such interactions can be indexed by coloured graphs $(\gamma,c)$ \cite{Bonzom:1202,Gurau:16}. 
In most of the literature one considers operators $\mathcal{O}_\gamma = \sum_c\mathcal{O}_{(\gamma,c)}$ which average over these colourings $c$ \cite{Benedetti:1411, Benedetti:1508, BenGeloun:1508, BenGeloun:1601, Carrozza:1612, Pithis:2007, Pithis:2009, BenGeloun:2305, Carrozza:1703, Carrozza:16, Lahoche:1904, Lahoche:2019}, reducing the phase space by a permutation symmetry. 
Here we consider the individual dynamics of each coloured operator $\mathcal{O}_{(\gamma,c)}$,
similar as done in \cite{BenGeloun:1201, Jepsen:2311} for tensor-valued scalar fields. 
This enriches the phase space drastically allowing for all the running of respective couplings $\lambda_{\gamma,c}$ under the renormalization group (RG) flow.

We calculate the RG flow using the non-perturbative method of the functional RG \cite{Berges:02, Delamotte:12, Dupuis:2021} in the cyclic-melonic potential approximation \cite{Carrozza:1612, Pithis:2007, Pithis:2009, BenGeloun:2305}.
This is the first order in a derivative expansion of the effective action,
here a local potential approximation (LPA) restricted to a specific consistent regime given by the most dominant tensorial interactions which are called melonic. 
For these interactions there is one type of graph $\gamma$ at each order $n$ of interaction with $\rk$ colourings $c$. 
For this reason we call the full phase space of couplings $\lambda_{n,c}$ ``anisotropic" while the symmetry-reduced one with all $\lambda_{n,c}=\lambda_n$ is called ``isotropic''.
We include also the running of the field renormalization parameter in which case the approximation is usually denoted as LPA$'$.

Applying for the first time such a LPA$'$, a first important result is that the anomalous dimension $\eta$ consists of contributions $\ec$ for each tensor field argument $c=1,...,\rk$.
We find that for a consistent RG flow it is necessary to assign individual field renormalization parameters $\wfc$ to each argument yielding also individual scale derivatives $\ec$.
Together with the mass parameter, these $\ec$ couple the different colour sectors.
Otherwise, the flow of couplings $\lambda_{n,c}$ depends only on couplings of the same colour $c$. 
As a consequence, we find classes of fixed points where only $s$ of the $c=1,...,\rk$ coloured couplings are non-zero. 
We call these partially interacting parts of phase space $s$-regimes.

In each such regime we find among the isotropic fixed points at least one non-Gaussian fixed point of Wilson-Fisher type, i.e.~with a single relevant direction in the $s$-subspace and smoothly connected to the Gaussian fixed point in critical dimension $\crd$. 
However, if $s\le\rk$, there are further relevant directions given by the scaling dimension in the direction of the vanishing couplings. 
Therefore, only the fixed point of the full $s=\rk$ regime has a single relevant direction providing a splitting of phase space in a broken and unbroken phase at small RG scale. 
For $s<\rk$, on the other hand, these fixed points have more than one but finitely many relevant directions and are thus asymptotic safe. 
Similar to previous results in the isotropic case \cite{Pithis:2009}, we find that all these fixed points exist even for the critical theory at $\rk-1=\crd=4$ due to the tensor-specific flow of the anomalous dimension.


Furthermore, we find a plethora of anisotropic non-Gaussian fixed points in each $s$-regime.
This is possible due to the indirect coupling of colour sectors via the quadratic flow equations.
We classify these fixed points at quartic truncation. 
Again, they occur not only for $\rk=4$ but also for the critical theory at $\rk=5$.

Overall, the remarkably rich phase space structure we classify opens up new possibilities to find interesting field-theoretical models both with respect to phase transitions when flowing to small RG scale as well as new asymptotic properties when flowing to large RG scale, e.g. asymptotic safe non-Gaussian fixed points.
Moreover, when tensorial interactions are interpreted as generating discrete geometries, the new fixed points open up the possibility to find quantum and random geometry built from subsets of polyhedral building blocks, and possibly new universality classes of continuum geometries.

\section{Anisotropic theory space}
\begin{figure}
\centering
\begin{tikzpicture}[x=5ex,y=5ex]
\begin{scope}
\node [v]       (v)     at (0,0)    {};
\foreach \i in {1,2}{
	\begin{scope}[rotate=\i *180]
	\node [wv]		(w\i)	at (-.6,.6)	{};
	\node [bv]		(b\i)	at (.6,.6)	{};
	\end{scope}
	}
\foreach \i in {1,2}{
	\path	(w\i) edge [eb] 		node 	{}	(b\i)
		(w\i)	edge [eb,bend left=30]node  {}	(b\i)
		(w\i)	edge [eb,bend right=30]node {}	(b\i)
		(w\i)   edge [e]            node    {}  (v)
		(b\i)   edge [e]            node    {}  (v);
	}
\foreach  \i/\j in {1/2,2/1}{
	\path (w\i) edge [eb] node {} (b\j);
	}
\node (c) at (-.4,0) {\scriptsize{$c$}};
\node (l) at (-1,0) {$\cn{2}^c$};
\end{scope}
\begin{scope}[xshift=15ex]
\node [v]       (v)     at (0,0)    {};
\foreach \i in {1,2,3}{
	\begin{scope}[rotate=\i *120]
	\node [wv]		(w\i)	at (-.5,.8)	{};
	\node [bv]		(b\i)	at (.5,.8)	{};
	\end{scope}
	}
\foreach \i in {1,2,3}{
	\path	(w\i) edge [eb] 		node 	{}	(b\i)
		(w\i)	edge [eb,bend left=30]node  {}	(b\i)
		(w\i)	edge [eb,bend right=30]node {}	(b\i)
		(w\i)   edge [e]            node    {}  (v)
		(b\i)   edge [e]            node    {}  (v);
	}
\foreach  \i/\j in {1/2,2/3,3/1}{
	\path (w\i) edge [eb] node {} (b\j);
	}
\node (c) at (-.6,.4) {\scriptsize{$c$}};
\node (l) at (-1.7,0) {,\quad$\cn{3}^c$};
\end{scope}
%
\begin{scope}[xshift=36ex]
\node [v]       (v)     at (0,0)    {};
\foreach \i in {1,2,3,4,5}{
	\begin{scope}[rotate=\i *63]
	\node [wv]		(w\i)	at (1.1,.3)	{};
	\node [bv]		(b\i)	at (1.1,-.3)	{};
	\end{scope}
	}
\foreach \i in {1,2,3,4,5}{
	\path	(w\i) edge [eb] 		node 	{}	(b\i)
		(w\i)	edge [eb,bend left=30]node  {}	(b\i)
		(w\i)	edge [eb,bend right=30]node {}	(b\i)
		(w\i)   edge [e]            node    {}  (v)
		(b\i)   edge [e]            node    {}  (v);
	}
\foreach  \i/\j in {1/2,2/3,3/4,4/5}{
	\path (w\i) edge [eb] node {} (b\j);
	}
\path	(w5) edge [eb, dashed, bend right=30]	node {} (b1);
\node (c) at (-1.2,.6) {\scriptsize{$c$}};
\node (s) at (-2.2,0)	{$,\dots, \;\cn{n}^c$};
\end{scope}
\end{tikzpicture}
\caption{The stranding of cyclic-melonic interactions is diagrammatically described by bipartite $\rk$-colourable vertex graphs (green edges, red vertices, $\rk=4$ in the example) with a distinguished edge colour $c\in\{1,2,...,\rk\}$. }\label{fig:cyclicmelonic}
\end{figure}
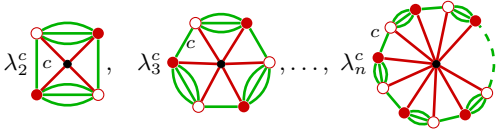

We consider a complex%
\footnote{Upon a simple rescaling \cite{Pithis:2009} one can derive the same results for real fields. Though this theory has a broader theory space of possible operators including e.g.~complete graphs playing a prominent role in melonic field theories \cite{Benedetti:2004}, these are suppressed at large $\ak$ \cite{Jepsen:2311} such that the theory space considered here is still closed and consistent.}
field $\phi: \gm^\rk \rightarrow \mathbb{C}$
of $\rk$ arguments valued in a group manifold $\gm$. For simplicity we restrict to one-dimensional $\gm$.
To treat tensorial interactions it is necessary to compactify to $\gm=U(1)\cong S_1$ with volume $\gv = \int_G\d g$. 
Specifically we consider as interactions all cyclic ($g_c^{n+1}=g_c^1$) melonic operators (Fig.~\ref{fig:cyclicmelonic})
\begin{equation}
        (\bar \phi \phi)_c^n = \int_{(\gm^\rk)^n} \prod_{i=1}^n \d\pmb{g}^i \ \bar\phi(g_1^i,..., g_r^i) \phi(g_1^i,..., g_c^{i+1},..., g_r^i) 
\end{equation}
for $n\ge2$ and $c=1,...,\rk$
which provide a closed theory subspace under the renormalization group flow in an appropriate limit \cite{Carrozza:1612}.
Here this is the limit of large $\ak = \gv k$ \cite{Benedetti:1411, Pithis:2009, BenGeloun:2305} which can be seen either as the limit removing the compactification $V\to\infty$ \cite{BenGeloun:1508, BenGeloun:1601} or for compact $G$ as the ``ultraviolet'' (UV) limit of large RG scale $k$. 
This scale is the momentum scale with respect to a bare kinetic term $\mathcal{K}=\sum_{c=1}^r \Delta_c + \mu$.
On $G=\U(1)$ this has spectrum $\sum_c (j_c/\gv)^\ks+\mu$ with $j_c\in\Z$.

The effective average action of the connected expectation values $\gf=\langle\phi\rangle_{\text{con}}$ in the cyclic-melonic regime is~\cite{Pithis:2009}
\begin{equation}
   \Gamma_k[\gf,\gfb] =  \int_{\gm^\rk} \!\!d\pmb{g} \ \gfb(\pmb{g}) \mathcal{K}_k 
   \gf(\pmb{g})  + \sum_{n\ge 2} \frac{1}{n!} \sum_{c=1}^r \cn{n,k}^c \left( \gfb \gf \right)^n_c \, .
\end{equation}
We find that for a consistent RG flow in the full aniso\-tropic phase space, i.e.~for independent couplings $\cn{n,k}^c$, it is \emph{necessary} to introduce a running field renormalization parameter $Z_{c,k}$ for each field argument $g_c$ individually,
\begin{equation}
    \mathcal{K}_k = \sum_{c=1}^r Z_{c,k} \Delta_c + \mu_k \, .
\end{equation}
%
We can still define a regulator fulfilling the standard conditions  of limit to effective action ($k\to 0$), limit to classical action ($k\to\Lambda$ UV scale), and positivity at $\pmb{j}^\ks\to 0$~\cite{Dupuis:2021}.
Explicitly, we use (again $\ak=\gv k$)
\begin{equation}\label{eq:regulator}
     \reg(\pmb{j}) 
     = \bigg( \wfr_k\k - \sum_{c=1}^r Z_{c,k} (j_c/\gv)^\ks\bigg)\, \theta\bigg( \ak^\ks - \sum_{c=1}^\rk j_c^\ks \bigg)
\end{equation}
as natural extension%
\footnote{
We have also checked the similar regulator with
$\wfc$'s also in the Heaviside function $\theta$. 
We use \eqref{eq:regulator} here as it yields simpler flow equations and satisfies the regulator conditions as well.
}
of the optimized regulator~\cite{Litim:2001ek} and 
\[
\wfr_k = \frac{1}{\rk}\sum_{c=1}^\rk Z_{c,k} \, .
\]
This satisfies the regulator conditions for all $\wfc\ge0$.

Given the regulator $\reg$, the Wetterich-Morris equation \cite{Wetterich:9302, Morris:9308} is a functional equation determining the RG flow of the effective average action $\Gamma_k$,
\begin{equation}\label{eq:FRGE}
    k\partial_{k} \Gamma_{k}[\gf,\bar{\gf}]=\frac{1}{2}\mathrm{Tr}\left[ k\partial_k \reg \left(\Gamma_{k}^{(2)}[\gf,\gfb]+\mathcal{R}_{k}\mathbb{I}_2\right)^{-1}  \right] \, ,
\end{equation}
for given initial condition $\Gamma_{k=\Lambda}$ at UV scale $\Lambda$.
Here, $\Gamma_{k}^{(2)}$ is the Hessian of $\Gamma_k$ and the trace $\mathrm{Tr}$ sums all degrees of freedom, here the domain $\gm^\rk$ and the two complex-field parts $\gf, \gfb$. 
Therein, the derivative
\begin{align}
k\partial_k \reg(\pmb{j}) 
&= \ks\wfr\k \left(1-\frac\ad\ks + \sum_c \frac\ec\ks \frac{j_c^\zc}{\ak^\ks} \right) 
\theta\bigg( \ak^\ks-\sum_{c=1}^\rk  j_c^\zc \bigg) \, 
\end{align}
involves the anomalous dimension $\ad$ as a scale derivative of $\wfr$ with contributions from each single $\wfc$,
\[
\ec = - \frac{k\partial_k \wfc}{\wfr} \, , \quad
\ad 
= - \frac{k\partial_k \wfr}{\wfr}
= \frac{1}{r} \sum_{c=1}^r \ec \, .
\]
This is a new property specific to tensorial interactions with independent couplings $\cn{n}^c$ for different colourings $c$.

\section{Anisotropic RG equations}

We derive the explicit form of the functional RG equation, \eqref{eq:FRGE}, using a vertex expansion, i.e.~comparing left- and right-hand side at each operator~$(\gfb\gf)^n_c$. 
To this end we expand the inverse of $\Gamma_{k}^{(2)}+\mathcal{R}_{k}\mathbb{I}_2$ in terms of a geometric series, the standard ``$PF$ expansion''~\cite{Benedetti:1411,BenGeloun:1601}.
This leads to non-autonomous equations in $\ak=\gv k$ which also include tensorial operators other than $(\gfb\gf)^n_c$.
However, in the autonomous, large-$\ak$ limit only these operators occur and one can rescale the couplings (suppressing explicit $k$ dependence from now on)
\begin{align}
    \mr &= \frac{\m}{\wfr\k} \, , \quad 
    \cnrc =  \ks I_{0}
    (\ak)^{n-1} \frac{\cn{n}^c}{(\wfr\k)^n} \,,
\end{align}
where $I_{0}$ are truncation-dependent spectral sums which always have large-$\ak$ asymptotics $I_{0}(\ak)\sim\ak^{\rk-1}$.
Thus the scaling is like that of a local scalar field theory in dimension \cite{Pithis:2009}
\[
\efd=\rk-1 \, .
\]
In particular, the critical theory is $\efd=\crd=4$, thus rank $\rk=5$ \cite{BenGeloun:1306}.

In this way we obtain for the first time the RG equations for all running parameters, $c=1,...,\rk\ge3$,%
\footnote{
In fact, \eqref{eq:flow equations} are valid also for $\rk=1$ in which case they agree exactly with those of vector theory in $\efd$ dimensions~\cite{Codello:1410}. 
The matrix case $\rk=2$ is special in that both $c=1,2$ interactions are not distinguishable such that there can be only a single coupling $\cnr{n}^1=\cnr{n}^2$. This yields specific factors in the beta functions indicating asymptotic safety \cite{Rivasseau:1507} 
}
$n\ge2$:
\begin{subequations}\label{eq:flow equations}
\begin{align}
    {\ec} &= \frac1\ks \bigg( \efd -\frac{
    3\ec - \ad}{\ks} \bigg) \bnc1
    \label{eq:eta-flow}\\
    k\partial_k\mr 
    &=(\ad-\ks)\mr + \sum_{c=1}^\rk \efac 
    \bnc1 
    \label{eq:mass-flow}\\
    k\partial_k\cnrc &=  -\efd\cnrc + n(\efd-\ks+\ad)\cnrc  
    + \efac \bnc{n} \,
    \label{eq:coupling flow}
\end{align}
\end{subequations}
with
$
\efac = 1 - \frac1{\ks} \frac{\ad+\ec}{\efd+\ks}
$
and $\bnc{n}$ are the beta functions  of large-$N$ vector theory (up to the linear, scaling part) with explicit form~\cite{BenGeloun:2305}
\[
\bnc{n} = \sum_{l=1}^n \frac{(-1)^l l!}{(1+\mr)^{l+1}} B_{n,l}(\cnr{2}^c,\cnr{3}^c,...,\cnr{n-l+2}^c) \, 
\]
in terms of partial exponential Bell polynomials~$B_{n,l}(x_1,...x_{n-l+1})=\sum_\sigma \binom{n}{s_1...s_n}\prod_{j=1}^{n-l+1}(x_j/j!)^{s_j}$ that sum over partitions $\sigma$ of $n$ with multiplicities $s_j$ for each part $j\in\sigma$ and length $\sum_j s_j =l$.

It is a specific property of melonic interactions that, at large $\ak$, the different colour sectors $c=1,...,\rk$ couple only through the quadratic part governed by $\m$ and $\ad$;
otherwise, the running of couplings $\cnrc$ depends directly only on couplings of the same colour $c$. 

This observation improves the understanding of the relation of melonic tensorial theories as compared to vector theories \cite{Pithis:2009}: concerning the interacting part, the tensorial theory simply behaves as $\rk$ independent copies of an $\O(N)$ vector theory at large $N$. 
That is, flow equations \eqref{eq:coupling flow} agree with those of large-$N$ vector theory for each $c=1,...,\rk$ individually.
However, at quadratic order, these otherwise independent dynamics couple through the the running of $\mr$ driven by all colour sectors equally.
Furthermore, all equations dependent on the full anomalous dimension $\ad$. 
This insight about melonic tensorial theories is in agreement with perturbative results \cite{Thurigen:2102,Thurigen:2103} where one also finds that 2-point correlations are built from tadpole insertions of any colour while 4-point correlations are series in ``bubble graphs'' (chains of fish graphs) of the same colour only.

This observation also explains the form of the RG equations in the isotropic sector.
In the \textit{isotropic limit} 
\begin{equation}
    \cnrc \rightarrow \cnr{n}
    \quad \text{for all } c=1,...,\rk \text{ and } n\ge2 \, ,
\end{equation}
\eqref{eq:flow equations} agree with the known flow equations \cite{BenGeloun:1601, Carrozza:1612, Pithis:2009}
as 
also $\ec=\ad$ 
due to \eqref{eq:eta-flow}.
In particular, the anisotropic equations explain the occurrence of the peculiar factor of $\rk$ in the isotropic $\mr$-flow equation~\cite{Pithis:2009}: it comes from the flow of $\mr$ receiving contributions from all colour sectors, given by the sum over all $c$ in \eqref{eq:mass-flow}.
On the other hand, the flow of couplings $\cnrc$ depend only on couplings of the same colour~$c$.

All the known isotropic fixed points \cite{Pithis:2009} are also solutions of the anisotropic flow equations
since the equations match in the isotropic limit.
That is, each fixed point $\cnr{n}=\cnr{n}^*$ also exists in the $\rk$-fold anisotropic phase space with $\cnrc=\cnr{n}^*$ for all $c=1,...,\rk$.
When truncating to $n \le \nmax$, this phase space has dimension $1+\rk (\nmax-1)$.
However, 
we find that,
except for the trivial case of the Gaussian fixed point,
the corresponding critical exponents do not simply occur with multiplicity $\rk$;
rather, each exponent of the isotropic phase space occurs once
as a ``radial'' exponent $\cer{n}$
in the anisotropic phase space and there is another new ``angular'' exponent $\cea{n}$ of multiplicity $\rk-1$. 
The Wilson-Fisher type fixed point in $\efd=3$, for example, has critical exponents in the LPA ($\ad=0$) at $\nmax=12$:
\[\label{eq:WP fixed point}
\theta_1 \approx 0.45217 \,,\,
\cea{2} \approx -0.38315 \,,\,
\cer{2} \approx -1.8761 \,, ... 
\]
where $\theta_1$ and $\cer{2}$ agree with Tab.~2 in \cite{Pithis:2009} while the three $\cea{2}$ are specific to the anisotropic phase space.

\

\section{Partially interacting fixed points}

\begin{table*}[ht]
\caption{\label{tab:r4exponents}%
$s$-isotropic non-Gaussian fixed point for $\rk=\efd+1=4$, $\eta=0$ in $(\gfb\gf)_c^{\nmax=12}$ truncation and their first exponents.
}
\begin{ruledtabular}
\begin{tabular}{rcccccccccccc}
$s$ & $\mr^*$ & $10\cnr{2}^*$ & $10^2\cnr{3}^*$ & $\theta_1$ &  $\ceg{2}$ & $\cea{2}$ & $\cer{2}$ & $\ceg{3}$ & $\cea{3}$ & $\cer{3}$ 
\\ 
\colrule
1 & -0.38835 & 2.9058 & 16.738
& 1.0000 & 1 & / & -1.0000 & 0 & / & -3.0000 \\
2 & -0.55944 & 1.0858 & 3.2449 
& 0.64994 & 1 & -0.38149 & -1.5265 & 0 & -2.6136 & -3.5635\\
3 & -0.65573 & 0.51811 & 0.94544 
& 0.51716 & 1 & -0.38149 & -1.7654 & 0 & -2.6136 & -3.9317 \\
4 & -0.71749 & 0.28633 & 0.35186 
& 0.45217 & / & -0.38149  & -1.8761 & / & -2.6136 & -4.1011
\end{tabular}
\end{ruledtabular}
\end{table*}

\begin{figure}
    \includegraphics[width=.9\linewidth]{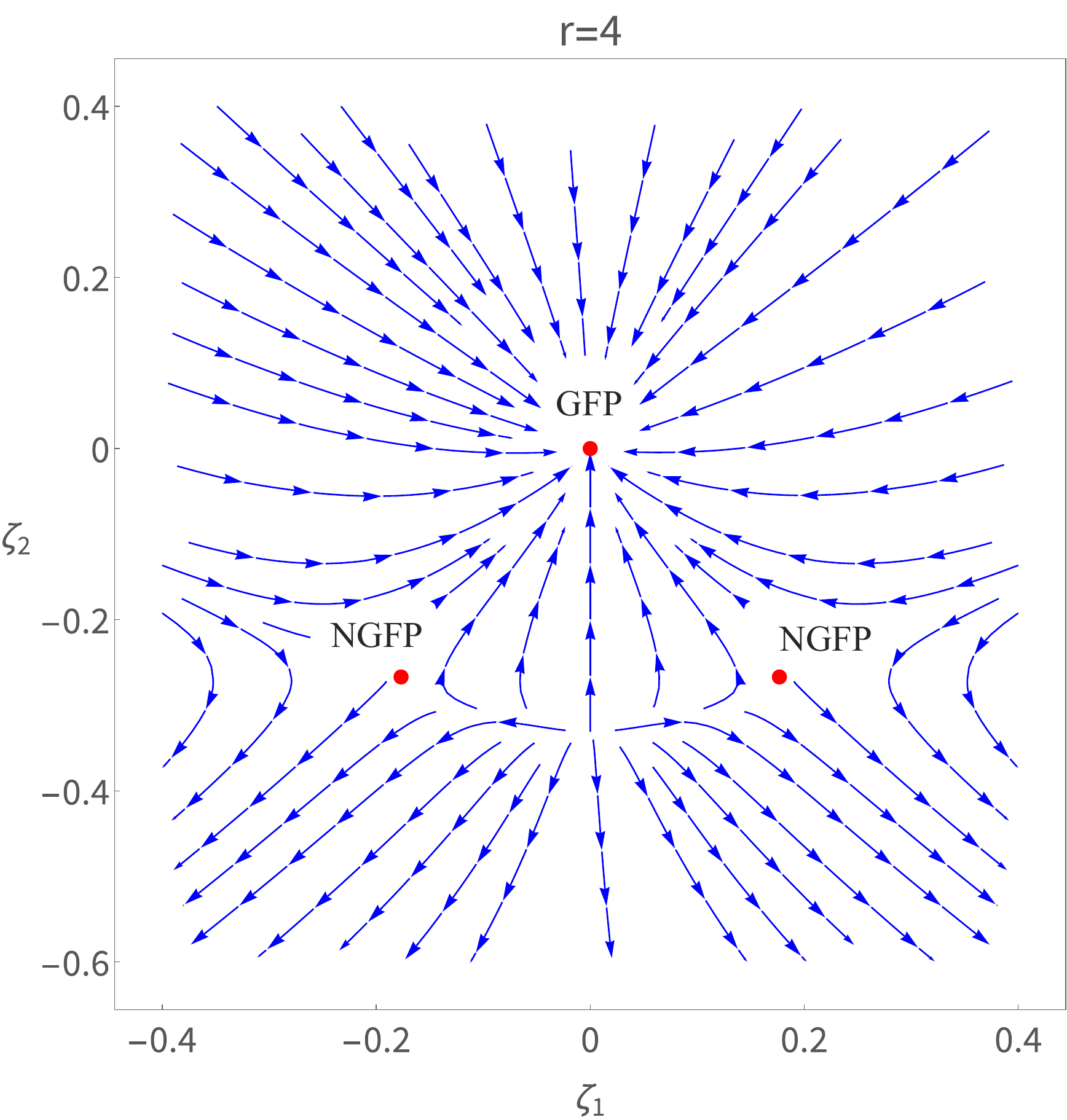}
    \caption{Flow diagram for the two $s=1$ isotropic fixed point for $r=4$, $\eta=0$ (Tab.~\ref{tab:r4exponents}) at quartic truncation when $\lambda_3 = \lambda_4=0$. 
    We choose a plane in $(\lambda_1, \lambda_2, \mu)$ space including both these non-Gaussian-fixed and the Gaussian fixed point ($\zeta_1$ and $\zeta_2$ are appropriate linear combinations of $\lambda_1$, $\lambda_2$ and $\mu$).} 
    \label{fig:flow1}
\end{figure}

An important consequence of the decoupling of colour sectors is that there are regimes in which interactions of some colours are turned off.
Consider initial conditions where, for all $n\ge2$, $\cnrc\ne 0$ only for $1\le c\le s<\rk$.
Then, according to \eqref{eq:coupling flow} there is no flow for 
$\cnr{n}^{s+1}=...=\cnr{n}^\rk=0$ and thus also 
$\eta_{s+1}=...=\eta_\rk=0$ due to \eqref{eq:eta-flow}.
Thus, the flow equations are those of only $s$ copies of the large-$N$ vector theory, though still with scaling dimensions of $\efd=\rk-1$.
In particular, the critical dimension remains $\efd =\crd=4$.
Nevertheless, only the subsector of $s$ out of $\rk$ types of interactions contributes to the dynamics.

In any such \emph{$s$-regime} there are again isotropic fixed points, $\cnrc=\cnr{n}$ and $\ec=\eta$ for all $c=1,...,s$ (and otherwise $\cnrc=\ec=0$)
such that $\ad=\frac{s}{\rk}\eta$.
In fact, taking into account permutations of colours $c$, there are $\binom{\rk}{s}$ such \emph{$s$-isotropic} fixed points for each $s\le\rk$ (see Fig.~\ref{fig:flow1}).
These are fixed-point solutions to the reduced flow equations
\begin{subequations}\label{eq:flow equations reduced}
\begin{align}
    \eta &= \frac1\ks \left( \efd -\left(
    3\rk/{s}-1\right)
    \frac{\ad}{\ks} \right) \bn1
    \label{eq:eta-flow_reduced}\\
    k\partial_k\mr &=\left(-\ks+\ad\right)\mr + s \efacs \, \bn1 
    \label{eq:mass-flow_reduced}\\[2pt]
    k\partial_k\cnrc &=  -\efd\cnrc + n\left(\efd -\ks +\ad\right)\cnrc  
    + \efacs \, \bn{n} \, ,
    \label{eq:coupling flow_reduced}
\end{align}
\end{subequations}
where now
\[
\efacs = 1 -\frac{
    \frac{\rk}{s}+1}{\efd+\ks}\frac\ad2 \, .
\]
They are very similar to the full isotropic equations \cite{Pithis:2009, BenGeloun:2305} except for a modification of the $\ad$ contribution to the flow by a fraction $\rk/s$ in some but not all terms.

\begin{table*}[ht]
\caption{\label{tab:r5exponents}%
The $s$-isotropic non-Gaussian fixed point for $\rk=\efd+1=5$ with $\eta^*<0$ in $(\gfb\gf)_c^{\nmax=10}$ truncation and exponents.
}
\begin{ruledtabular}
\begin{tabular}{rcccccccccccc}
$s$ & $\mr^*$ & $\ad^*$ & $10\cnr{2}^*$ & $10^2\cnr{3}^*$ & $\theta_1$ &  $\ceg{2}$ & $\cea{2}$ & $\cer{2}$ & $\ceg{3}$ & $\cea{3}$ & $\cer{3}$ 
\\ 
\colrule
1 &  -0.38160 & -0.13109 & 2.9186 & 16.530
& 0.78152 
& 0.26217 & / & -2.4369 
& -1.6067 & / & -3.4757 \\
1 & -0.47879 & -0.23236 & 2.6014 & 15.580
& 0.79536
& 0.46472 & / & -2.4250
& -1.3029 & / & -3.0706 \\
1 & -0.55347 & -0.39334 & 2.2071 & 13.091
& 1.0362
& 0.78669 & / & -1.9567
&-0.81997 & / & -9.3463 \\
2 & -0.28277 & -0.11869 & 1.5374 & 2.2712 
& 1.0565
& 0.23739 & -0.74750 & -3.3581 
& -1.6439 & -1.2248  & -3.5003 \\
2 & -0.62457 & -0.45014 & 1.0357 & 3.2407
& -0.32774
& 0.90029  & -2.2771 & -5.4425
& -0.64957 & -3.8345 & -7.3678 \\
3 & -0.72775 & -0.69033 & 0.45113 & 0.88234
& 1.4936 
& 0.54809 & 1.3807 & -1.6337 
& -1.2387 & 0.071000 & -4.9059\\
3 & -0.87521 & -1.0897 & 0.12352 & 0.19475  
& 2.0970 & 2.1794 & 1.5334 & -1.9772 & 1.2691 & 0.35881 & -14.163 \\
3 & -0.96742 & -1.5470 & 0.0099619 & 0.0057458
& -2.3200
& 3.0941 & -0.52954 & -37.022
& 2.6411 & -6.0245  & -106.83 \\
4 & -0.95561 & -1.3584 & 0.013206 & 0.0071868 
& -0.61347
& 2.7168 & -2.4397 & -15.942
& 2.0752 & -22.734 & -49.439 \\
5 & -0.94714 & -1.4194 & 0.014634 & 0.0071649
& 0.54992 
& / & -0.56823 & -8.1386 
& / &  -12.171 & -28.966\\
\end{tabular}
\end{ruledtabular}
\end{table*}

\

In the LPA, neglecting the anomalous dimension $\ad$, one recovers the fixed-point values of rank-$s$ tensor theory on $\efd$-dimensional space \cite{BenGeloun:2305}.
Factors $\rk/s$ do not occur and the parameter $s$ is independent of $\efd=\rk-1$, though restricted here to an integer $s\le\rk$.
However, as $s$-isotropic fixed points in anisotropic phase space their critical exponents have a specific pattern (Tab.~\ref{tab:r4exponents}):
like in the $s=\rk$ case, \eqref{eq:WP fixed point}, there is one radial exponent $\cer{n}$ and now $s-1$ angular exponents $\cea{n}$ for each $n\ge2$ which we find to be independent of $s$.
In particular, for the vector analogue $s=1$ we recover the exponents
\[
\cer{n}=\efd-\ks n
\]
of the $\efd$-dimensional Wilson-Fisher fixed point which are understood analytically \cite{Tetradis:9512,DAttanasio:9704}.
On the other hand, the $\rk-s$ zero couplings $\cnrc=0$ each contribute simply an exponent which is the scaling dimension,
\[\label{eq:scaling exponents}
\ceg{n}=\efd - n(\efd -\ks) \, ,
\]
like for the Gaussian fixed point. 
That is, in an $s$-regime of anisotropic phase space non-Gaussian fixed points still have partial Gaussian behaviour.

An immediate consequence is that there exist asymptotic safe fixed points in the full cyclic-melonic LPA for $\rk=4$.
All the $s$-isotropic fixed points (Tab.~\ref{tab:r4exponents}) converge for large $\nmax$ and are of Wilson-Fisher type in the sense that their radial exponents are negative, $\cer{n}<0$, for $n\ge2$ but $\theta_1>0$.
Yet, for $s<\rk$ there are also Gaussian directions that add further relevant directions. 
This gives also the explanation for the number of relevant directions for the respective fixed points in \cite{Jepsen:2311}.
Overall, there is only one fixed point with a single relevant direction, the $\rk$-isotropic one. 
This allows for phase transitions between a broken and unbroken phase for $k\to 0$ \cite{Pithis:2007}.
All other fixed points (as we will explain in the following) have more than one relevant directions.
For the $s$-isotropic ones these are finitely many, that is $1+\rk-s$ relevant directions.

\

\begin{figure}
    \includegraphics[width=.9\linewidth]{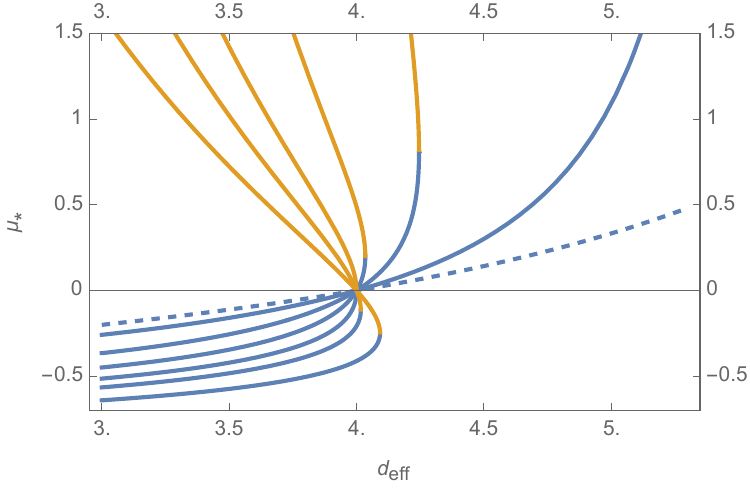}
    \caption{Fixed point solutions $\mr^*$ at quartic truncation as function of $\efd=\rk-1$ of the $s$-sector equations \eqref{eq:flow equations reduced} for 
    $s = 1, 1.5, 2, 2.5, 3, 4$ 
    from right to left (at $\mr_*>0$); 
    for comparison, the dashed curve is the $\efd$-dimensional large-$N$ (thus $\eta=0$) Wilson-Fisher fixed-point solution.}
    \label{fig:mass-s-isotropic}
\end{figure}

In the LPA$'$ (running $\eta$) we find branchings of these fixed points.
To illustrate this, we show in Fig.~\ref{fig:mass-s-isotropic} the fixed-point solutions $\mr^*$ as functions of $\efd$.
They are of Wilson-Fisher type in the sense that they coincide with the Gaussian fixed point at the critical dimension $\efd=\crd=4$.
For $s=1$, this is still a single curve with a pole at some $\efd>\crd$. 
Around $s\approx 1.625$ this pole turns into a branching point such that for all larger $s$ the curve has two branches; that is, there are two real non-Gaussian fixed points up to some $\efd=\nosed>\crd$.
While for $s/\rk<1/2$ both branches have positive mass, $\mr_*(\efd)>0$ for $\efd>\crd$, these mass solutions are negative for $s/\rk<1/2$.
For $s=\rk$ this agrees with previous results~\cite{Pithis:2009}.

An immediate consequence is that there are Wilson-Fisher type fixed points even at criticality $\rk=\crd+1=5$ for $s>5/2$.
However, only for $s=5$, this fixed point (agreeing with \cite{Pithis:2009}, Tab.~3) has a single relevant direction;
for $s<\rk$, there are again $(\rk-s)$-fold exponents from the, now $\eta$-dependent, scalings
\[\label{eq:scaling exponents2}
\ceg{n}=\efd - n(\efd -\ks + \ad) \, ,
\]
such that all these $s$-isotropic fixed points provide again \emph{asymptotic safe} fixed points.
Moreover, we find that at higher-order truncations there occur even more branchings of the solution curve smoothly connected to the critical Gaussian fixed point.
We show converging fixed points of this type (in particular $\ad^*<0$) at order $\nmax=10$ truncation in Tab.~\ref{tab:r5exponents}.

\section*{Anisotropic fixed points}

\begin{figure}
    \includegraphics[width=.9\linewidth]{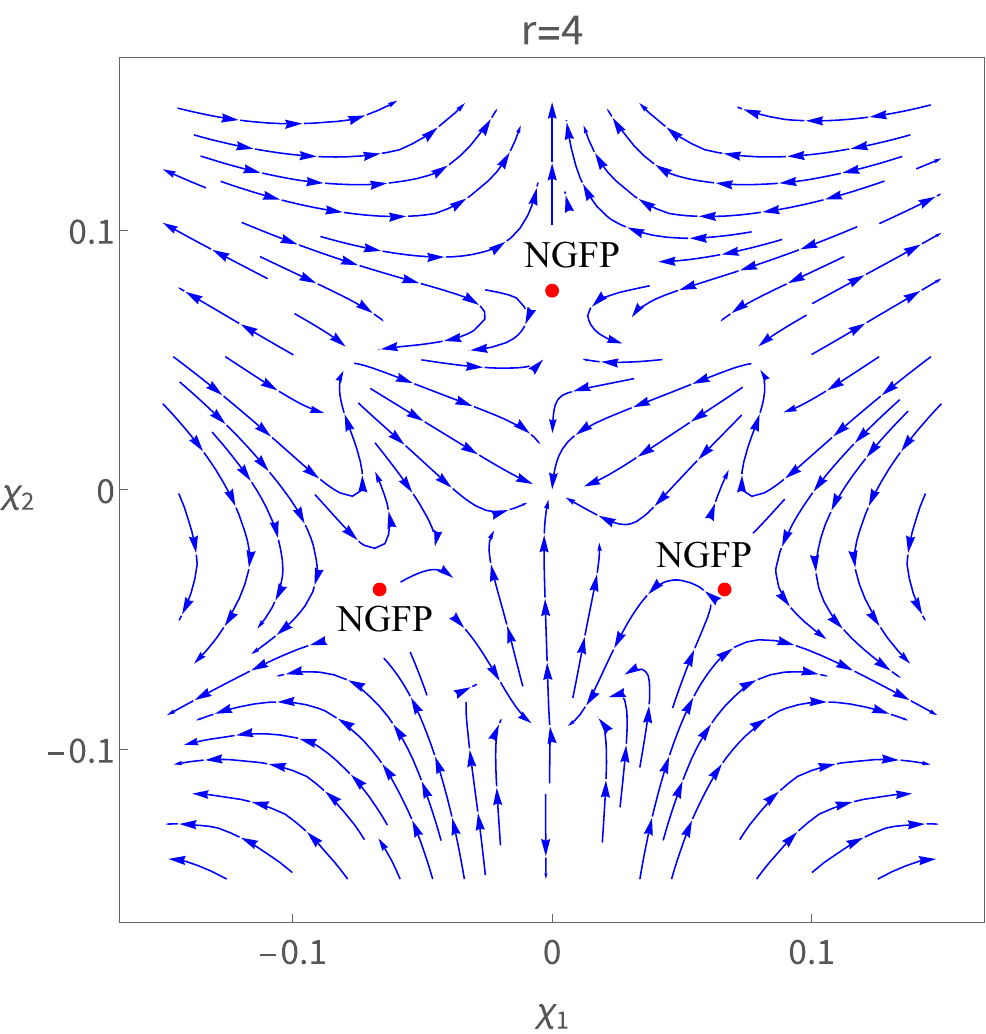}
    \caption{Flow diagram for three permutations of the single $(2,2)$ type fixed point in $r=4$ with running $\eta$ at quartic truncation. 
    Fixing $\mu = -0.86 $ and $\lambda_4 = -0.043 $ we show a linear slicing of the $(\lambda_1, \lambda_2, \lambda_3)$ parameter space including these three non-Gaussian-fixed points.} 
    \label{fig:flow4}
\end{figure}

The anisotropic RG flow equations have also fixed-point solutions which are not $s$-isotropic.
Though the colour sectors decouple, the indirect coupling through $\mr$ and $\ad$ allows for anisotropic solutions.
In an $s$-regime with non-vanishing couplings of $s$ colours, solutions of any partition $(s_1,s_2,...)$, $\sum_j j s_j =s$ with respective permutations of the colours are possible in principle (see Fig.~\ref{fig:flow4} for an example).
This means, a fixed point of type $(s_1,s_2,s_3...)$ has 
$s_1$ couplings of value $\cnr{n}^{c}=\cnr{n}^1$,
$s_2$ couplings $\cnr{n}^{c'}=\cnr{n}^2$ etc.~.
However, at quartic-order truncation, only fixed points of double type $(s_1,s_2)$ occur as we explain in the Appendix.
With higher order, more types of anisotropic fixed points are possible.

For the $(s_1,s_2)$-type fixed points we can use again reduced flow equations to explicitly determine their values and exponents.
Define 
$\eta^1 = \frac{s_1}{\rk}\eta_c$ and
$\eta^2 = \frac{s_2}{\rk}\eta_{c'}$
for all $c=c_1,...,c_{s_1}$ and $c'=c_{s_1+1},...,c_{s_1+s_2}$.
Then the flow equations simplify for $(a,b)=(1,2),(2,1)$ to
\begin{subequations}\label{eq:flow equations reduced2}
\begin{align}
    \eta^a &= \frac{s_a}{\ks\rk} \!\left(\efd - (
    3\rk/s_a-1)\frac{\eta^a}\ks + \frac{\eta^b}\ks
     \right) \beta_1(\mr,\cnr{2}^a)
    \label{eq:eta-flow_reduced2}\\
    k\partial_k\mr &=\left(-\ks+\eta^1+\eta^2\right)\mr 
    + \!\!\!\sum_{a\ne b=1,2}
     \!\! s_a \efacsa \beta_1(\mr,\cnr{2}^a) 
    \label{eq:mass-flow_reduced2}\\
    k\partial_k\cnr{n}^a &=  -\efd\cnr{n}^a + n\left(\efd-\ks +\eta^1+\eta^2 \right)\cnr{n}^a 
    \label{eq:coupling flow_reduced2}\\
    & \quad\quad\quad\quad\quad\quad\quad\quad\quad + \efacsa \, \beta_n(\mr,\cnr{i}^a)
    \nonumber
\end{align}
\end{subequations}
with
\[
\efacsa =  1 -\frac{
(\frac{\rk}{s_a}+1)\frac{\eta^a}\ks + \frac{\eta^b}\ks}{\efd+\ks} \, .
\]
These flow equations have only the complexity of a rank-two theory and can thus be solved up to higher-order truncations when the full anisotropic equations, \eqref{eq:flow equations}, already become unfeasible. 
At quartic truncation, there are four solutions except for the special types $(s-1,1)$ with three solutions and $(\frac{s}{2},\frac{s}{2})$ with two solutions (see the Appendix).
We present results for $\efd = \crd$ in Tab.~\ref{tab:r5exponents12}.



\begin{table*}[ht]
\caption{\label{tab:r5exponents12}%
$(s_1,s_2)$ type fixed points for rank $\rk=\crd+1=5$ with running $\eta$ at quartic truncation.}

\begin{ruledtabular}
\begin{tabular}{r ccccc  cccc}
$(s_1, s_2)$ & $\mr^*$ & $\eta_c^* = \frac\rk{s_1}\eta^{1*}$ &  $\eta_c^* = \frac\rk{s_2}\eta^{2*}$ & $\cnr{2}^{*} \ (s_1)$ & $\cnr{2}^{*} \ (s_2)$ & $\theta_1$ &  $\ceg{2}$ & $\cea{2}$ & $\cer{2}$
\\ 
\colrule
(4, 1) & -0.87423 & 5.3773 & 7.5628 & 0.33816 & 0.063274
& 87.055
& / & 14.251 & -6.6718 \\
(4, 1) & -1.6667 & -1.3333 & 8.6667 & 0.18714 & 0.88889
& 9.6107  
& / & 1.3333 & -0.27747     \\
(4, 1) & -7.6258 & -14.877  & 4.1871 & 49.287 & 175.60
& 6.1464 
& / & 0.74846 & -0.029879   \\
(3, 2) & -0.88956 & 2.4321 & 9.5360 & -0.019852 & 0.030340  
& 64.833
& / & 10.547 & -4.8224  \\
(3, 2) & -0.87656 & 4.4527 & 8.2973 & -0.19367 & 0.049911 
& 82.660
& / & 13.519 & -6.3059\\
(3, 2) & -3.0869 & -3.6520 & 6.6195 & 3.2774 & 10.114   
& 7.1174
& / & 0.91319 & -0.094887  \\
(3, 2) &-4.6161 & -6.9684 & 5.3235 &12.434 & 36.812
& 6.5293
& / & 0.81353  & -0.054897   \\
(2, 2) & -0.86633 & 2.6650 & 10.396  & -0.036422 & 0.041362  
& 53.711 
& -10.449 & 10.449 & -4.6924    \\
(2, 2) & -3.0781 &-4.7312 & 6.0824 & 3.5949 & 10.824
& 6.9731 
& -1.0809 & 1.0809 & -0.10887    \\
(3, 1) & -0.85024 &4.6088 &9.7456 & -0.43859 & 0.060429
& 69.156
& -13.538& 13.538  & -6.2342 \\
(3, 1) & -1.5728 & -1.4594 & 8.4486 & 0.14518 & 0.67068
& 9.6752
& -1.6282 & 1.6282 & -0.34842  \\
(3, 1) & -5.3177 & -11.328  &4.5606 & 19.897 & 65.263 
& 6.2905
& -0.94166 & 0.94166 & -0.053016  \\
(2, 1) & -0.82707 & 3.1416 & 11.844  & -0.095352 & 0.063919
& 44.279
& -10.732 & 10.732 & -4.7174  \\
(2, 1) & -1.4656 & -1.5647 & 8.2495 & 0.098904 & 0.45492
&9.6148
& -2.0481& 2.0481 & -0.44651 \\
(2, 1) & -3.2999 & -7.3170 & 5.2762 & 5.0597 & 15.545 
& 6.6440
& -1.2941 & 1.2941 & -0.11458  \\
(1, 1) & -1.3333 & 8.1867  & -1.5201 & 0.23896 & 0.048626
& 9.2436
& -2.6667 & / & -0.57697   \\
\end{tabular}
\end{ruledtabular}
\end{table*}

\section{Conclusions}

Here we have generalized the functional RG equations for tensorial field theory of arbitrary rank $\rk$ in the cyclic-melonic potential approximation from the isotropic \cite{Pithis:2009} to the full anisotropic case.
To this end it proves necessary to introduce independent field renormalization parameters $\wfc$ for each field argument $g_c$ resulting in individual contributions $\ec$ to the anomalous dimension $\ad$.

We find that the structure of the RG equations is that of $\rk$ copies of the vector-theory equations but coupled at quadratic order.
This improves the understanding of the relation of vector and tensor theories.
As a consequence there exist partially interacting regimes with only $s\le\rk$ colour types contributing, similar to results in \cite{Jepsen:2311}. Accordingly, such fixed points have both Gaussian and interacting properties which are manifest in the structure of critical exponents. 
We classify both isotropic and truly anisotropic fixed points in each such $s$-regime.

Despite the plethora of newly found fixed points, only the $\rk$-isotropic non-Gaussian fixed point has a single relevant direction. 
Thus, the general result of a phase transition between a broken and unbroken phase at small scale~$k$, previously found in the restricted isotropic case~\cite{Pithis:2007}, remains in the full anisotropic phase space.
Due to the partially Gaussian behaviour of non-Gaussian fixed points in $s<\rk$ regimes, all other such fixed points have more than one relevant directions.
However, at least for the $s$-isotropic fixed points of Wilson-Fisher type in the interacting directions we can show that there are also only finitely many relevant directions, thus providing a classification of asymptotic safe non-Gaussian fixed points.

We have restricted ourselves here to the regime of cyclic-melonic interactions. 
It is straightforward to broaden the phase space further including more types of tensor interactions. 
Inclusion of multi-trace or simple-graph interactions as common in tensor field theories will enlarge the phase space to some extent, like in \cite{Jepsen:2311}, though there are no different colour types for these interactions.
Much more additional structure is expected when including for example so-called necklace interactions whose isotropic phase space has already been investigated \cite{Carrozza:1703} since these have even $\binom{\rk}{\rk/2}$ colour variants per interaction type. 

One important application of tensor theories is random and quantum geometry \cite{Rivasseau:1112, Gurau:16, Oriti:1110}.
In such an interpretation, the tensorial theory is a generating function of $\rk$-dimensional triangulations with a continuum limit in terms of the non-perturbative, critical regime.
In the dynamical picture of the renormalization group this is exactly the non-Gaussian fixed points.
It is known in tensor models that the interplay of differently coloured interactions provides a plethora of (multi)critical regimes, typically in the universality classes of Liouville quantum gravity, branched polymers and baby universes \cite{Lionni:1707}.
In this sense our results provide the RG dynamics to which continuum geometry phase the theory is flowing for given initial conditions.

Within this interpretation, we find that there exist continuum-geometry limits generated from various different classes of triangulations.
First of all, with the $s$-regimes there are continuum geometries from triangulations made of certain subclasses of building blocks given by the interactions. 
Then, we see that there are not only geometries where all types of buildings blocks contribute equally, given by the $s$-isotropic fixed points;
but there are also continuum limits where the various types contribute with a certain ratio as provided by the values of anisotropic fixed points. 
Thus, it will be a crucial task for future research to analyse the geometric properties of the continuum limits given by the new non-Gaussian fixed points we have presented here.

\section*{Acknowledgements}
J.~Th\"{u}rigen's research was funded by DFG grant number 418838388 and Germany's Excellence Strategy EXC 2044--390685587, Mathematics M\"unster: Dynamics–Geometry–Structure. 

\section*{Appendix}
We provide here a simple characterization of the fixed points to quartic truncation, recalling that there is evidence to discard points with $|\mu|>1$ as they usually diverge when raising the order of truncation. 
The idea is to solve \eqref{eq:flow equations} analytically for $\partial_k \Tilde{\mu}_k = \partial_k \Tilde{\lambda}a_k=0$ performing the change of variables 
\begin{equation}
    \Tilde{\lambda}^c_2 \rightarrow l^c_2 = \frac{4\Tilde{\lambda}^c_2}{4(1+\mu)^2-3\Tilde{\lambda}^c_2}
\end{equation}
and using the $F$ function in the $\eta$ equation to get an expression of $\mu$ as a function of $l^c$. After simple algebraic manipulations it is possible to recast $\mu$ and the $\eta^a$'s as functions of $l^c$:
\begin{equation}
    \begin{split}
        \eta^i & = l^i \left( \frac{1-r}{2}-\frac{(1-r)L}{8r + 2L}\right) \equiv l^i a(L)\\
        \mu(L,s) & = \frac{s (3 Lr+L-4 (r-5) r)}{r (L (3 s+2)+20 s+8)+L s-4 r^2 s}
    \end{split}
\end{equation}
where $L:= \sum_c l^c$ and $s$ is, again, the number of non-zero $l^c$'s. These two expressions can be plugged back into the $\Tilde{\lambda}^c_2$ equation, to get: 
\begin{equation}
    (l^i)^2 a(L) + l^i b(L) - c(L) = 0  \qquad \forall \ i \in \ 1, \dots, s
\end{equation}
where again $b(L)$ and $c(L)$ are appropriate functions of $L$. The solution is: 
\begin{equation} \label{appendixBfieq}
    l^i_{\pm} = \frac{-b\pm\sqrt{b^2+4ac}}{2a}
\end{equation}
Calling $s_1$ the number of $+$ solutions and $s_2$ the number of $-$ solutions we can sum over $i$ to obtain an equation for $L$ only: 
\begin{equation}  \label{appendixBFeq}
     2La(L) = -b(L) + (s_1-s_2) \sqrt{b^2+4ac}
\end{equation}
These equations allow to extract all the qualitative features of the structure of fixed points to quartic truncation: 
\begin{itemize}
    \item since \eqref{appendixBfieq} is second order, we can have only $(s_1, s_2)$ sectors;
    \item as \eqref{appendixBFeq} is of fourth order, the anisotropic fixed points can be at most four; 
    \item imposing $s_1 =s_2$ the equation reduces to second order;
    \item when $s_1 = 1$ and $s_2 \ne 1$ the equation reduces to third order;
    \item if $s_1=s_2=1$ the equation becomes of first order. 
\end{itemize}
All these features are manifest in Tab.~\ref{tab:r5exponents12}.

\bibliographystyle{apsrev4-2}
\bibliography{main.bib}

\end{document}